\documentstyle[aps,prl,preprint,floats,epsfig]{revtex}

\begin{document}

\newcommand{\lbc}{$\Lambda_{c}^{+}$}

\preprint{\tighten\vbox{\hbox{\hfil CLNS 00/1701}
                        \hbox{\hfil CLEO 00-24}
}}

\title{Measurement of the \lbc~Lifetime}  

% Your author list ***DOES NOT*** go here!
% is goes below where you are instructed to insert it...
\author{CLEO Collaboration}
\date{\today}

\maketitle
\tighten

\begin{abstract} 
The \lbc~lifetime is measured  
using 9.0~fb$^{-1}$ of $e^+e^-$ annihilation data collected
on or just below the $\Upsilon(4S)$ resonance with the CLEO II.V detector
at CESR.
Using an unbinned maximum likelihood fit, the \lbc~lifetime is
measured to be $(179.6 \pm 6.9 ({\rm stat.}) \pm 4.4 ({\rm syst.}) )$ fs.
The precision of this colliding beam measurement 
is comparable to other measurements, which are based on
fixed-target experiments, with different systematic uncertainties. 

\end{abstract}
\newpage

{
\renewcommand{\thefootnote}{\fnsymbol{footnote}}

% Insert author and address list here

\begin{center}
A.~H.~Mahmood,$^{1}$
S.~E.~Csorna,$^{2}$ I.~Danko,$^{2}$ K.~W.~McLean,$^{2}$
Sz.~M\'arka,$^{2}$ Z.~Xu,$^{2}$
R.~Godang,$^{3}$ K.~Kinoshita,$^{3,}$%
\footnote{Permanent address: University of Cincinnati, Cincinnati, OH 45221}
I.~C.~Lai,$^{3}$ S.~Schrenk,$^{3}$
% ...
\end{center}
 
\small
\begin{center}
$^{1}${University of Texas - Pan American, Edinburg, TX 78539}\\
$^{2}${Vanderbilt University, Nashville, Tennessee 37235}\\
$^{3}${Virginia Polytechnic Institute and State University,
Blacksburg, Virginia 24061}\\
% ...
\end{center}

\setcounter{footnote}{0}
}
\newpage

% Insert body of the text here.

Lifetime measurements of heavy quark mesons and baryons
provide an important window into the non-perturbative sector of heavy quark decay.
Contrary to initial expectations~\cite{theory}, mechanisms other than spectator
quark decay make significant contributions to the lifetimes of
weakly decaying charm mesons and baryons. 
Charm baryon lifetimes differ by large amounts 
({\em e.g.} $\tau_{\Xi_{c}^{+}}$:$\tau_{\Lambda_{c}^{+}}$:$\tau_{\Xi_{c}^{0}}$~$\sim$~
4:2:1)~\cite{pdg}, as is also seen in the charm mesons.
However, the underlying reasons for this pattern may be different than
for the mesons as $W$ exchange in baryon decay is neither helicity nor
color-suppressed.  Other effects, such as Pauli interference, may
also play a different role in the baryon sector~\cite{theory}.   
This paper reports a new measurement of the lifetime of the
\lbc, the lowest-mass charm baryon, with a precision
comparable to that from measurements of the charm meson lifetimes.   
The data used in this analysis were obtained in an $e^+e^-$ colliding beam environment,
where the event topologies and backgrounds are very different from those
encountered in high energy fixed-target experiments~\cite{fixed-target},
which have historically provided the most precise measurements of charm hadron
lifetimes~\cite{pdg}.

This analysis uses an integrated luminosity
of 9.0 fb$^{-1}$ of $e^+e^-$ annihilation data recorded with the 
CLEO~II.V detector at the Cornell Electron Storage Ring (CESR). 
The data were taken at energies at or slightly below the $\Upsilon(4S)$ resonance 
($\sqrt{s}=10.58$ GeV) and 
contain approximately 11 million $e^+ e^- \to c\bar{c}$ events.
The CLEO II.V detector upgrade consists of an interaction
region (IR) and a change in the gas used in the primary tracking volume
(refer to~\cite{cleoii} for a detector description before the upgrade).  The
IR consists of a small-radius, low-mass beam pipe surrounded by a three-layer
double-sided silicon vertex detector (SVX).  The SVX records precision 
3-dimensional tracking information close to the interaction point~\cite{beampipe,svx}.  
The proximity of
the SVX to the IR, combined with the low mass beam-pipe, delivered excellent vertex resolution.  
The momentum resolution was enhanced as a result of replacing
the primary tracking volume gas from a 50:50 mixture of argon-ethane to a 60:40
mixture of helium-propane.  This change increased the hit efficiency and 
decreased the effects of multiple scattering.
The helium-propane replacement also enhanced specific ionization information
used for particle identification.  Further, to optimize the data from the upgraded
detector, a Kalman filter track reconstruction package~\cite{kalman} was implemented.
The response of the CLEO detector 
to both signal and background events was simulated
in detail using a GEANT-based~\cite{geant} Monte Carlo package.
 
The \lbc~is reconstructed in the $pK^{-}\pi^{+}$ decay mode 
(the charge conjugate mode is implied throughout this paper).
General track quality and event shape cuts are used to remove poorly reconstructed
tracks and non-hadronic events.   
A decay vertex measurement is needed for a proper time measurement, so we require 
at least two of the three decay tracks to
have 2 or more hits simultaneously on a track in both the $xy$ and $rz$~\cite{coordinates} views.
The efficiency to have two or more SVX hits simultaneously
in both views is 95\% per track, yielding 99\% efficiency per \lbc;
the average decay vertex resolution was about 110~$\mu $m.

The combinatoric background to the \lbc~signal is suppressed by taking advantage
of the detector upgrades.
Tracks forming a \lbc~candidate are required to originate from a common vertex
(${\rm x_{dec}}, {\rm y_{dec}}, {\rm z_{dec}}$)
in a 3-dimensional vertex fit. We require the vertex fit probability to be $>0.001$. 
Particle identification information from specific ionization 
must  be consistent with the \lbc~daughter hypothesis. 
Electrons are rejected using drift chamber and calorimeter information.  
Backgrounds tend to populate the low track momentum spectrum and 
low \lbc~momentum spectrum.
Each decay track is therefore required to have momentum greater than 0.3~GeV/$c$, and
the \lbc~momentum is required to be greater than
2.6~GeV/$c$.  As a result of these cuts, the selected \lbc's
have an average momentum of 3.3~GeV$/c$. 

Combinatoric backgrounds from random combinations of tracks from 
$D^{0}$, $D^{+}$, and $D^{+}_{s}$ decays, 
which may populate the reconstructed mass region non-uniformly
and bias the lifetime result, are studied using data and Monte Carlo.   
From tests using CLEO data and Monte Carlo simulated
events, we find that $D^{+} \rightarrow K^{-}\pi^{+}\pi^{+}$ events
(where one of the pions passes the proton requirements) preferentially populate the
reconstructed mass region above the \lbc~mass peak.  
Such events are a potentially problematic kinematic reflection that we
remove by rejecting candidates whose reconstructed mass is 
consistent with a $D^{+}$ mass when the proton hypothesis
is changed to a pion hypothesis.
All other backgrounds ({\em e.g.} $D^{+}_{s} \rightarrow \phi \pi^{+}$) are
found to be either uniform throughout the mass region or small enough
not to affect the final lifetime result.   

\begin{figure}[htb]
\centering
\epsfig{figure=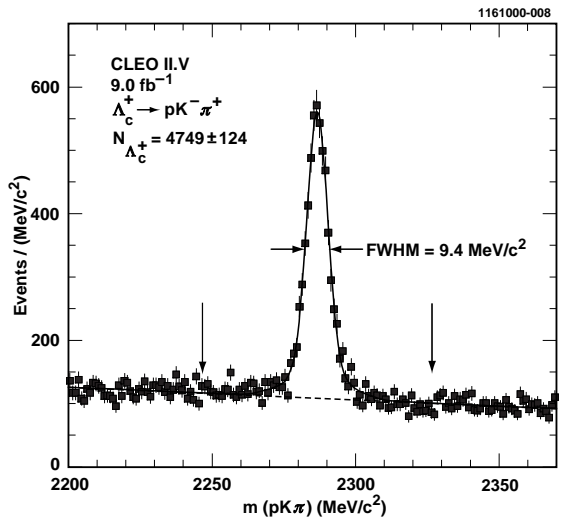,width=3.0in}
\caption{\label{fig:mass}
Reconstructed mass distribution of \lbc~candidates.
The data (squares) are fit 
(solid line) to two Gaussians constrained to a single mean and a linear background (dotted line).
There are 4749 $\pm$ 124 $\Lambda_{c}^{+} \rightarrow pK^{-}\pi^{+}$
signal events.
All events between the arrows ($\pm 40$~MeV/$c^2$, or $\pm$ 
11.1 $\sigma_{\rm{narrow}}$, of the   
\lbc~signal peak) are used in the nominal unbinned maximum likelihood fit.
} 
\end{figure}

The reconstructed mass distribution of the \lbc~candidates is shown in
Fig.~\ref{fig:mass}, and a fit to the data
yields 4749 $\pm$ 124 $\Lambda_{c}^{+} \rightarrow pK^{-}\pi^{+}$ 
signal events.
The mass distribution is fit using two Gaussians ($\sigma_{\rm{narrow}}$, $\sigma_{\rm{broad}}$)
constrained to a common mean for the signal and a linear function for the background. 
The fraction of background in the mass region within 7~MeV/$c^2$ 
(1.94 $\sigma_{\rm{narrow}}$) of the fitted \lbc~mass
value is 27.2\%, while  
85.2\% of the \lbc's are within this region as well.  

\begin{figure}[htb]
\centering
\epsfig{figure=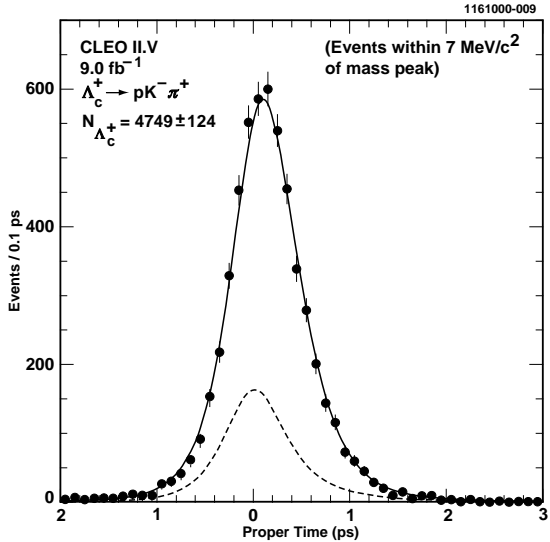,width=3.0in}
\caption{\label{fig:propt}
Reconstructed proper time distribution (points) of \lbc~candidates 
within 7~MeV/$c^2$ of the \lbc~signal peak.
Overlaid is a representation of the
unbinned maximum likelihood lifetime fit scaled
to the $\pm 7$~MeV/$c^2$ reconstructed mass region (solid line).
The fitted background component scaled
to the $\pm 7$~MeV/$c^2$ reconstructed mass region is also shown (dotted line).}
\end{figure}

The \lbc~production point (${\rm x_{prod}}, {\rm y_{prod}}, {\rm z_{prod}}$) is needed 
for the proper time determination. 
A run-averaged beam centroid determined from two-track and hadronic events
provides an estimate of the \lbc~production point.  
The dimensions of the beam profile are about 350~$\mu $m, 7~$\mu $m, and 1~cm along the $x$,
$y$, and $z$ directions, respectively, as determined by CESR~\cite{beam-spot} optics.  
Given the extent of the profile in $x$ and $z$, the
measurement is effectively determined by the $y$ component. 
The $y$-components of the decay length, 
$l_{\rm y} \equiv {\rm y_{dec}} - {\rm y_{prod}}$, and \lbc~
momentum, $p_{{y}_{\Lambda_{c}^{+}}}$, are used to determine the proper decay times, 
calculated from
$t = m_{\Lambda_{c}^{+}} l_{\rm y}/c p_{{y}_{\Lambda_{c}^{+}}}$ 
using the PDG~\cite{pdg} world average for $m_{\Lambda_{c}^{+}}$. 
The decay length variance $\sigma^{2}_{t}$ is calculated using the results of the
vertex fit and the beam profile dimension uncertainties. 
The proper time distribution for the \lbc~candidates
within 7~MeV/$c^2$ of the \lbc~signal peak 
is shown in Fig.~\ref{fig:propt}.  The average decay length is about
80 $\mu$m, comparable to the vertex resolution.
 
The \lbc~lifetime is extracted from the proper time distribution with
an unbinned maximum likelihood fit~\cite{dmesonlife}.  
There are three inputs to the fit for each \lbc~candidate:  the measured 
proper time $t_{i}$, the estimated uncertainty of the measured proper time $\sigma_{t,i}$, and a
reconstructed mass-dependent
signal probability $p_{{\rm sig},i}$.  The signal probability distribution is obtained from 
a fit of the reconstructed mass distribution to two Gaussians constrained to the same mean
for the signal and a linear function for the background.
The likelihood function is

\widetext
\begin{eqnarray}\label{eq:likelihood}
  L&&(\tau_{\Lambda_{c}^{+}}, 
   f_{\rm bg}, 
  \tau_{\rm bg}, 
  S, 
  f_{\rm mis}, 
  \sigma_{\rm mis}, 
  f_{\rm wide} ) \nonumber\\
  &&=\prod_i \int_0^\infty dt^\prime 
  \left[
    \underbrace{p_{{\rm sig},i}E(t^\prime|\tau_{\Lambda_{c}^{+}})}_{\rm signal\ fraction} +
    \underbrace{(1-p_{{\rm sig},i})\left[ f_{\rm bg} E(t^\prime|\tau_{\rm bg})
    + (1- f_{\rm bg})\delta(t^\prime)\right]}_{\rm background\ fraction}
  \right] \nonumber \\
  &&\ \ \ \times \left[
    \underbrace{(1-f_{\rm mis}-f_{\rm wide})G(t_i-t^\prime|S
    \sigma_{t,i})}_{\rm proper~time\ resolution} +
       \underbrace{
         f_{\rm mis} G(t_i-t^\prime|\sigma_{\rm mis}) +
         f_{\rm wide}G(t_i-t^\prime|\sigma_{\rm wide})}_{\rm mismeasured\ fraction}
  \right],
  \nonumber
\end{eqnarray}
\narrowtext

\noindent where the product is over the \lbc~candidates, 
$G(t|\sigma)\equiv \exp(-t^2/2\sigma^2)/\sqrt{2\pi}\sigma$,
and $E(t|\tau) \equiv \exp(-t/\tau)/\tau$.
The seven output parameters of the lifetime fit
are $\tau_{\Lambda_{c}^{+}}$,  $f_{\rm bg}$, $\tau_{\rm bg}$, $S$, 
$f_{\rm mis}$, $\sigma_{\rm mis}$ and $f_{\rm wide}$. 
The parameter $\tau_{\Lambda_{c}^{+}}$ is 
the \lbc~lifetime.  Each candidate is weighted in the fit according to its proper time
uncertainty $\sigma_{t,i}$.  The fit allows for a global scale factor $S$
for the proper time uncertainty estimates.  For a small fraction of 
candidates $f_{\rm mis}$ the fitted uncertainty $S \sigma_{t,i}$ 
underestimates the true uncertainty.  Track 
reconstruction errors such as those caused by hard multiple scattering
are examples of such mismeasurements.  
The proper time distribution of the background is modeled by a fraction $f_{\rm bg}$ having a
background lifetime $\tau_{\rm bg}$ with the remaining background having zero
lifetime. In order to estimate the background properties, we include
the candidates in a wide region of $\pm 40$~MeV/$c^2$ 
($\pm$ 11.1 $\sigma_{\rm{narrow}}$) around the nominal
\lbc~mass in the unbinned maximum likelihood fit. 
We account for mismeasured candidates with two Gaussians 
in the fit.  In order to accommodate a small fraction $f_{\rm wide}$
of candidates that fall outside the resolution parameterization, a 
wide Gaussian ($\sigma_{\rm wide}$ = 8 ps) is used to approximate a flat distribution.   
Fig.~\ref{fig:propt} shows the proper time distribution of events
within $7$~MeV/$c^2$
of the \lbc~signal peak and the results of the unbinned maximum
likelihood fit scaled to the $\pm~7$~MeV/$c^2$ reconstructed mass
region.  The fit converges to $\tau_{\Lambda_{c}^{+}} = (178.6 \pm 6.9)$~fs,
where the uncertainty is statistical only, and 
the correlations between the \lbc~lifetime and the other fit
parameters range from $-0.26$ to 0.10. 

Consistency checks are performed 
by measuring the lifetime as functions of azimuthal angle, polar angle,
momentum of the \lbc~candidate, charge of the
candidate, and data taking period. 
No statistically significant variation is found in any of these variables.
The lifetime is also measured as a function of reconstructed mass in
the background regions for CLEO data and simulated Monte
Carlo events.  The lifetimes are 
consistent with being uniform, 
and thus properly parameterized.  

\begin{table}[htb]
  \begin{center}
  \caption{Contributions to the systematic uncertainty for 
           the \lbc~lifetime.}
    \label{tab:syst-tot-all}
    \medskip  
    \begin{tabular}{|l|c|}
      Contribution & Uncertainty (fs)  \\ \hline
      Decay vertex resolution & $\pm 2.2$ \\
      Global detector scale & $\pm 0.1$ \\
      Beam spot position    & negligible \\
      \lbc~mass measurement & $\pm 0.05$ \\
      \lbc~momentum measurement & $^{+0.2}_{-0.1}$ \\
      Signal probability from mass fit & $^{+0.5}_{-0.7}$ \\
      $t$ -- $m(pK\pi)$ correlation & $\pm 2.3$ \\
      Large proper times & $\pm 0.9$ \\
      Statistics of Monte Carlo sample & $\pm 2.8$ \\
      \hline
      Total &  $\pm 4.4$ \\ 
    \end{tabular}
  \end{center}
\end{table}

The contributions considered in the systematic uncertainty for the \lbc~ 
lifetime are listed in
Table~\ref{tab:syst-tot-all} and are described below.
Decay vertex measurement errors lead to decay length errors which in turn
could lead to an error in the lifetime measurement.   
A $(0.0 \pm 0.9)~\mu $m bias in the decay vertex position is estimated
from a ``zero-lifetime'' sample of 
$\gamma\gamma \to \pi^+\pi^-\pi^+\pi^-$ events. 
To obtain the corresponding proper time bias,
the error of the decay vertex position bias is multiplied
by the average 1/$\gamma \beta c$ of the 
\lbc candidates.  The vertices
of events having interactions at the beam pipe are used to determine
a relative radial position uncertainty of $\pm 0.2$\%.  The quadrature sum of these uncertainties 
yields the systematic uncertainty due to the decay vertex measurement.   
The global detector scale is known to a precision of $\pm 0.1$\% 
from detailed surveys using beam pipe interactions.  
A systematic uncertainty of $\pm0.1$~fs is assigned.
In order to determine the sensitivity of the analysis to the beam spot position, 
the vertical beam spot position is shifted $\pm 2~\mu $m, the uncertainty of the 
beam spot position, 
and the decay lengths of the \lbc~candidates from CLEO data are recalculated.  
The differences between these shifted lifetimes and the nominal lifetime
are included as a systematic uncertainty.

The uncertainty in the \lbc~mass~\cite{pdg} 
and the \lbc~momentum scale 
lead to systematic errors since these quantities are used in the conversion of 
decay lengths to proper times.  
The statistical uncertainty of the signal probability assigned to each 
\lbc~candidate leads to systematic uncertainties in the fitted lifetimes, which are  
estimated by coherently varying the signal
probability of each candidate by its statistical uncertainty and repeating
the fits.   
There also exists a correlation between the measurements of the proper time $t$ and the
\lbc~candidate reconstructed mass $m_{pK\pi}$.
This correlation is measured in simulated events and confirmed 
in data to estimate a corresponding systematic uncertainty.

Poorly measured \lbc~candidates are another source of 
systematic uncertainty. These candidates are accounted for in the 
unbinned maximum likelihood fit with a wide Gaussian that
approximates a flat distribution ($\sigma_{\rm wide}$ = 8 ps).  Another method of
accounting for candidates with large proper times is to omit the wide Gaussian
component from the likelihood function and fit the candidates 
in a restricted proper time interval.  The systematic uncertainty due to these
candidates is estimated from the maximum variation of
$\tau_{\Lambda_{c}^{+}}$ in varying the width of the wide Gaussian from 8 to
12 ps and removing the wide Gaussian in a restricted proper time fit
($\mid t \mid<4$~ps).

Other possible sources of lifetime measurement bias include \lbc~
selection requirements and parameterization of the background in the  
likelihood function.  These are checked by performing the unbinned maximum likelihood
fit on an artificial sample composed of background events extracted from 
$e^+e^-$ annihilation data and simulated 
$\Lambda_{c}^{+} \rightarrow pK^{-}\pi^{+}$ events with known average lifetime of 206.2 fs. 
Accurately measuring the lifetime of this combined sample is an important test of the 
unbinned maximum likelihood method.  
The lifetime obtained from the
maximum likelihood fit, $(205.2 \pm 2.8)$~fs, is consistent with the known input lifetime 
of the signal Monte Carlo sample. 
The statistical uncertainty of this measured lifetime,
2.8 fs, is included as a contribution to the total systematic uncertainty. 
The lifetime difference of $-1.0$ fs is subtracted from the lifetime extracted from 
data yielding a corrected \lbc~lifetime of $(179.6 \pm 6.9)$~fs, where the
uncertainty is statistical only.  
The total systematic uncertainty of 4.4 fs is obtained by adding the individual
contributions in quadrature. 

In summary, a new measurement of the \lbc~lifetime
using 9.0~fb$^{-1}$ of integrated luminosity has been made with 
the CLEO II.V detector.  The measured \lbc~lifetime is
$(179.6 \pm 6.9 \pm 4.4)$~fs,
where the first uncertainty is statistical and the
second systematic.  This is the first \lbc~lifetime measurement
made in a non-fixed target environment.

We gratefully acknowledge the effort of the CESR staff in providing us with
excellent luminosity and running conditions.
This work was supported by 
the National Science Foundation,
the U.S. Department of Energy,
the Research Corporation,
the Natural Sciences and Engineering Research Council of Canada, 
the A.P. Sloan Foundation, 
the Swiss National Science Foundation, 
the Texas Advanced Research Program,
and the Alexander von Humboldt Stiftung.


\begin{thebibliography}{99}

\bibitem{theory}
G.~Bellini, I.~Bigi, and P.J.~Dornan, Phys. Rept. {\bf 289}, 1 (1997) and 
references contained within.

%\relax
\bibitem{pdg}
Particle Data Group, D.~E.~Groom {\em et al.}, Eur.~Phys.~J.~C~{\bf 15}, (2000). 

\bibitem{fixed-target} 
E687 Collaboration,  P.L.~Frabetti {\em et al.}, Phys. Rev. Lett. {\bf 71}, 
827~(1993);\\ % Ds 
Phys. Lett.~{\bf B 323}, 459~(1994);\\ % D0, D+
E691 Collaboration, J.R.~Raab {\em et al.}, Phys. Rev. D~{\bf 37},
2391~(1988);\\  %  D0, D+, Ds % \relax
E791 Collaboration, E.M. Aitala {\em et al.}, Phys. Lett.~{\bf B 445},
449~(1999). % Ds

\bibitem{cleoii} 
CLEO Collaboration, Y. Kubota {\em et al.}, Nucl. Instrum. Methods A {\bf
  320}, 66 (1992). 

\bibitem{beampipe}
The inner radius of the beam pipe is 1.875 cm.  The three layers of silicon 
are located at 2.35, 3.25, and 4.75 cm from the detector origin.

\bibitem{svx}
T. Hill, Nucl. Instrum. Methods A {\bf 418}, 32 (1998). 
%\relax

\bibitem{kalman}
P.~Billoir, Nucl. Instrum. Methods A {\bf 255}, 352~(1984).

\bibitem{geant}
  We use a GEANT-based computer model to simulate the response of the CLEO 
  detector.
  R. Brun {\em et al.}, GEANT 3.15, CERN Report No.~DD/EE/84-1 (1987). % \relax

\bibitem{coordinates}
The right handed coordinate system has the $z$ axis along the $e^+$ beam
direction and the $y$ axis upward.

\bibitem{beam-spot}         
  D.~Cinabro {\em et al.}, Phys. Rev. E {\bf 57}, 1193 (1998). % \relax

\bibitem{dmesonlife}
  G.~Bonvicini {\em et al.}, Phys. Rev. Lett.~{\bf 82}, 4586~(1999). 

\end{thebibliography}
\end{document}